\documentclass[reprint,preprintnumbers,twocolumn,
superscriptaddress,
amsmath,amssymb,aps]{revtex4-2}
\usepackage[utf8]{inputenc}
\usepackage{mathrsfs}

\usepackage{graphicx}
\graphicspath{{./figs/}}
\usepackage{dcolumn}
\usepackage{bm}

\usepackage{enumitem}

\usepackage{array}

\usepackage{color}
\usepackage[dvipsnames, table]{xcolor}
\definecolor{blue}{rgb}{0,0,0.5}
\definecolor{lightgray}{gray}{0.95} 

\usepackage[colorlinks,linkcolor=blue,urlcolor=blue,citecolor=blue]{hyperref}
\usepackage{slashed}
\usepackage{physics}
\usepackage{xspace}
\usepackage[T1]{fontenc}
\usepackage{lmodern}
   
\usepackage{tabularx} 
\usepackage{booktabs}  
\usepackage{diagbox}
\usepackage{multirow}

\newcommand{\be}{\begin{equation}}
\newcommand{\ee}{\end{equation}}
\newcommand{\bea}{\begin{eqnarray}}
\newcommand{\eea}{\end{eqnarray}}
\newcommand{\<}{\langle}
\renewcommand{\>}{\rangle}
\newcommand{\mc}{\mathcal}

\newcommand{\nn}{\nonumber}
\newcommand{\noi}{\noindent}
\def\bk{\vb*{k}}

\def\min{{\rm min}}
\def\dof{{\rm dof}}
\def\hc{{\rm h.c.}}

\def\inv{{\rm inv}}
\def\diag{{\rm diag}}

\newcommand{\MeV}{{\rm MeV}}
\newcommand{\GeV}{{\rm GeV}}

\newcommand{\SM}{{\rm SM}}
\newcommand{\BR}{{\mc B}}
\newcommand{\Kppipa}{K^+ \to \pi^+ a}

\def\refeq#1{Eq.~(\ref{#1})}
\def\refeqs#1#2{Eqs.~(\ref{#1})-(\ref{#2})}
\def\reftab#1{Table~\ref{#1}}
\def\reffig#1{Fig.~\ref{#1}}
\def\reffigs#1#2{Figs.~\ref{#1}-\ref{#2}}
\def\refsec#1{Sec.~\ref{#1}}

\def\refcite#1{Ref.~\cite{#1}}

\def\nunu{\nu \bar \nu}
\def\de{\partial}

\def\Ms{\mc M_{s}}
\def\Mw{\mc M_{w}}

\def\Laa{\mathscr L_a}
\def\Las{\mathscr L_s}
\def\Law{\mathscr L_w}
\def\FV#1{(F_{V})_{#1}}
\def\FA#1{(F_{A})_{#1}}

\newcommand{\PQ}{{\rm PQ}}
\newcommand{\muPQ}{\mu_{\PQ}}
\newcommand{\muEW}{\mu_{\rm EW}}
\newcommand{\muK}{\mu_K}

\def\fab{\bar f_a}
\def\az{|z|}
\def\azexpl{|1 + \boldsymbol{\alpha}\cdot\mathbf{c}|}

\def\adotc{\boldsymbol{\alpha}\cdot\mathbf{c}}

\def\Unif{\mathrm{Unif}}

\newcommand{\camalich}{MartinCamalich:2020dfe}

\newcommand{\mael}{Cavan-Piton:2024ayu}
\newcommand{\neubertPRL}{Bauer:2021wjo}
\newcommand{\neubertALPslong}{Bauer:2021mvw}
\newcommand{\neubertRGE}{Bauer:2020jbp}

\newcommand{\georgi}{Georgi:1986df}

\newcommand{\ecker}{Ecker:1987qi}
\newcommand{\diluzio}{DiLuzio:2020wdo}
\newcommand{\cornella}{Cornella:2023kjq}
\newcommand{\isidori}{Isidori:2005xm}

\newcommand{\beal}{\begin{aligned}}
\newcommand{\eeal}{\end{aligned}}

\makeatletter
\g@addto@macro\bfseries{\boldmath}
\makeatother

\makeatletter
\def\p@subsection{}
\makeatother

\newlist{todolist}{itemize}{2}
\setlist[todolist]{label=$\square$}
\usepackage{pifont}
\DeclareOldFontCommand{\rm}{\normalfont\rmfamily}{\mathrm}
\DeclareOldFontCommand{\sf}{\normalfont\sffamily}{\mathsf}
\DeclareOldFontCommand{\tt}{\normalfont\ttfamily}{\mathtt}
\DeclareOldFontCommand{\bf}{\normalfont\bfseries}{\mathbf}
\DeclareOldFontCommand{\it}{\normalfont\itshape}{\mathit}
\DeclareOldFontCommand{\sl}{\normalfont\slshape}{\@nomath\sl}
\DeclareOldFontCommand{\sc}{\normalfont\scshape}{\@nomath\sc}

\begin{document}

\preprint{LAPTH-008/25}

\title{New bound on the vectorial axion-down-strange coupling\\from $K^+ \to \pi^+ \nu \bar \nu$ data}

\author{Diego Guadagnoli}

\email{diego.guadagnoli@lapth.cnrs.fr}

\affiliation{%
{\itshape LAPTh, Universit\'{e} Savoie Mont-Blanc et CNRS, 74941 Annecy, France}
}%

\author{Axel Iohner}

\email{axel.iohner@lapth.cnrs.fr}

\affiliation{%
{\itshape LAPTh, Universit\'{e} Savoie Mont-Blanc et CNRS, 74941 Annecy, France}
}%

\author{Cristina Lazzeroni}

\email{cristina.lazzeroni@cern.ch}

\affiliation{%
{\itshape University of Birmingham, United Kingdom}
}%

\author{Diego~Martínez~Santos}

\email{diego.martinez.santos@cern.ch}

\affiliation{%
{\itshape Ferrol Industrial Campus, Dr. Vázquez Cabrera, s/n, 15403, Universidade de A Coruña, Spain}
}%

\author{Joel C. Swallow}

\email{joel.swallow@cern.ch}

\affiliation{%
{\itshape CERN, Esplanade des Particules 1, 1211 Meyrin, Switzerland}
}%

\author{Claudio Toni}

\email{claudio.toni@lapth.cnrs.fr}

\affiliation{%
{\itshape LAPTh, Universit\'{e} Savoie Mont-Blanc et CNRS, 74941 Annecy, France}
}%

\begin{abstract}
\noi We reinterpret publicly available $K^+ \to \pi^+ \nu \bar{\nu}$ data collected by NA62 from 2016 to 2024 to constrain the fundamental vectorial coupling of the QCD axion to down and strange quarks. Using a fully reproducible likelihood analysis and a complete renormalization-group evolution of the axion couplings from the Peccei–Quinn (PQ) scale to the kaon scale, we translate the experimental limit into bounds on both the low-energy flavour-violating coupling and the fundamental UV parameters. In the generic regime where strong contributions dominate the decay amplitude, we obtain $|\FV{sd}(\mu_K)| > 1.6 \times 10^{12}\,\GeV$. 
The coexistence of parametrically suppressed weak contributions implies a second, conceptually distinct constraint: the fact that weak-amplitude dominance arises from highly tuned UV coupling configurations, translates into a conservative general lower limit on the PQ scale, $f_a > 4.9 \times 10^4\,\GeV$.
These results provide the strongest accelerator-based constraints on axion-induced $d \leftrightarrow s$ transitions and establish a robust lower bound on $f_a$, complementary to astrophysical limits.
\end{abstract}

\maketitle

\noi {\bf Introduction---} %
New spin-zero, odd-parity particles below the GeV scale---collectively referred to as axions or axion-like particles (ALPs)---are hypothetical states, searched for by an expanding range of laboratory experiments~\cite{Irastorza:2018dyq,Lanfranchi:2020crw,Sikivie:2020zpn}, astrophysical observations~\cite{Raffelt:1990yz,Raffelt:2006cw,Caputo:2024oqc}, cosmological probes~\cite{Marsh:2015xka}, and phenomenological considerations, see \emph{e.g.}~\cite{Kim:2008hd,DiLuzio:2020wdo,Choi:2020rgn}.

A well-defined thread within this broader context is the search for the so-called ``invisible'' \cite{Kim:1979if,Shifman:1979if,Dine:1981rt,Zhitnitsky:1980tq} QCD axion~\cite{Peccei:1977hh,Peccei:1977ur,Weinberg:1977ma,Wilczek:1977pj}, hereafter simply {\em the axion}. This is a highly plausible particle, as it provides a uniquely elegant resolution to the strong-$CP$ problem, and a compelling framework for accounting for the fraction of the universe's total density attributed to Dark Matter~\cite{Preskill:1982cy,Abbott:1982af,Dine:1982ah}. Unlike generic ALPs, the axion’s mass $m_a$ and decay constant $f_a$ are intrinsically related: smaller values of $m_a$ correspond to weaker interactions with standard matter, with couplings suppressed parametrically as $1/f_a$. Current constraints imply that $m_a \ll m_\pi$~\cite{DiLuzio:2020wdo}. Despite the accordingly feeble couplings, the small mass means that axions may be produced, potentially even copiously, within controlled accelerator environments, without the need to push the energy frontier.

Among the many experimental approaches to probing the existence of axions, one of the most direct is their potential production in decays recorded at high-intensity facilities such as NA62. This approach is well motivated for several reasons.
First, this method has for decades been used to study decays to---and even to unambiguously identify~\cite{Cowan:1956rrn}---neutrinos, now-established feebly interacting, uncharged, near-massless particles with properties largely analogous to those of axions. Second, from a theoretical point of view {\em light} meson decays involving an axion admit a systematic, effective-theory description, known as Chiral Perturbation Theory (ChPT) \cite{Gasser:1983yg,Weinberg:1978kz}, which ensures reliable predictions with a controlled uncertainty. Axion interactions with light mesons can be included in this description by a careful generalization~\cite{\georgi} of the same logic that leads to the construction of ChPT. In this construction, starting from the Lagrangian terms
\be
\label{eq:Laqq}
\Laa ~\supset~
\frac{\de_\mu a}{2 f_a} 
\left(
\bar q \, \gamma^\mu \bk_{V} \, q + 
\bar q \, \gamma^\mu \gamma_5 \bk_{A} \, q
\right) - \frac{\alpha_s}{8 \pi} \frac{a}{f_a} G \tilde G~,
\ee
with $q = (u, d, s)^T$~\footnote{Our definition of the Peccei–Quinn scale coincides with that in the review~\cite{\diluzio}. The sign difference in the defining $a G \tilde G$ term in \refeq{eq:Laqq} arises from our use of opposite $\epsilon$-tensor conventions relative to those in Ref.~\cite{\diluzio}; see Supplemental~\refsec{app:ChPT}.}, the ChPT couplings to hadrons are unambiguously determined once the $\bk_{V,A}$ coupling matrices are fixed---up to the standard hadronic low-energy constants at the chosen chiral order and to an unphysical reparametrization ambiguity~\cite{\georgi,\neubertPRL}.

The axion-down-strange couplings $(k_{V,A})_{sd}$---the only allowed off-diagonal entries in \refeq{eq:Laqq}---are thus two fundamental couplings of the QCD axion to matter. Given that they generate $d \leftrightarrow s$ transitions otherwise highly suppressed in the SM, these couplings can lead to ``very dramatic'' effects \cite{\georgi}.
The strongest probe of the vectorial $|(k_V)_{sd}|$ are $K \to \pi a$ decays. The diagonal axial couplings $(k_A)_{uu,dd}$ (reals) are best probed by Neutron Stars~\cite{Buschmann:2021juv}. Finally, it has recently been shown that Supernova cooling induced by axions radiated from strange matter in the star core~\cite{\mael,Fischer:2024ivh} is sensitive to the axial $|(k_A)_{sd}|$ and $(k_A)_{ss}$.

Here we probe $K^+ \to \pi^+ a$ by reinterpreting experimental studies of the ultra-rare decay $K^+ \to \pi^+ \nu \bar \nu$, which provides one of the most exquisite precision tests of the Standard Model (SM). The tiny $O(10^{-10})$ branching fraction of the $K^+$ into this decay mode \cite{Buras:2022wpw,Anzivino:2023bhp,DAmbrosio:2022kvb} is dominated by short-distance contributions, and the necessary hadronic matrix element can be extracted from {\em data} using isospin. The decay is thus uniquely rare and uniquely clean theoretically. The NA62 experiment at CERN was designed with the measurement of $\BR(K^+ \to \pi^+ \nu \bar \nu) \equiv \BR_{\nunu}$ as its primary goal. Using the decay-in-flight technique, it has reported a sequence of increasingly precise measurements based on the 2016–2018 data~\cite{NA62:2018ctf,NA62:2020fhy,NA62:2021zjw}, the 2021–2022 dataset~\cite{NA62:2024pjp}, the full 2016–2022 sample~\cite{NA62:2025upx}, and most recently the 2023–2024 data~\cite{LaThuile_2026_update}.

Using {\em all} these measurements, one can constrain decays to $\pi^+$ plus new invisible states, in particular $K^{+}\rightarrow\pi^{+}a$. Such a signal would appear as an excess over $K^{+}\rightarrow\pi^{+}\nu \bar \nu$. At an intuitive level, the experimental uncertainty $\delta \BR_{\nunu}$ quantifies the available room for this excess and thus sets the expected sensitivity to $\BR(K^{+}\rightarrow\pi^{+}a)$. The combination of 2016–2024 NA62 data yields $\delta \BR_{\nunu}\simeq 1.9\times10^{-11}$, predominantly statistical. Our first result is a fully reproducible likelihood-based reinterpretation relying exclusively on public data, which quantitatively confirms this expectation.

We apply this method to infer the strongest existing bound on $|(k_V)_{sd}|/f_a$ at the kaon scale---our second result, obtained from the entire 2016–2024 dataset analyzed so far. This result will remain state-of-the-art until the 2025-2026 data, currently being collected by NA62, are analyzed. Using a full renormalization-group evolution from the Peccei–Quinn scale to hadronic scales, we express this constraint directly in terms of the fundamental UV couplings and the PQ scale $f_a$. This allows us to expose unambiguously the hierarchy between the strong and parametrically suppressed weak contributions to the amplitude. Accordingly, two conceptually distinct bounds emerge. In the generic regime of strong-amplitude dominance we obtain the leading constraint. Conversely, weak-amplitude dominance requires highly tuned UV coupling configurations. This tuning requirement allows the weak contribution to define a conservative general floor on the PQ scale, extracted from controlled accelerator data. This forms our third result.

\noi {\bf Reinterpretation Technique---} %
To bound the $K^+ \to \pi^+ a$ decay through the $K^+ \to \pi^+ \nu \bar \nu$ measurements at NA62, we build on the procedure of Ref.~\cite{NA62:2020xlg}.
We perform a fully frequentist hypothesis test using a shape analysis of the invisible invariant mass and an unbinned profile likelihood ratio test statistic, based solely on public data. The likelihood function is constructed as the product of three contributions, i.e. ${\cal L}={\cal L}_1 {\cal L}_2 {\cal L}_3$.

The first term is a Poisson distribution for the total number of events $n_\text{tot}$ with expectation value fixed to the total number of observed events $n_\text{obs}$.
The variable $n_\text{tot}$ can be decomposed as $n_\text{tot}=n_b+n_a$, with $n_{b,a}$ being the number of background and axion events, respectively.

The number $n_a$ is related to the BR as $\BR(\Kppipa) = n_a\times\BR_{\rm SES}$, where $\BR_{\rm SES}$ denotes the associated single-event sensitivity (SES). $\BR_{\rm SES} = \BR_{\rm SES}(m_a)$ is provided in \emph{e.g.} the rightmost panel of Figure 2 of Ref.~\cite{NA62:2020xlg}, which pertains to the 2017 data. 
Limits for other datasets are obtained by rescaling with the ratio of the corresponding $K^+\to\pi^+\nu\bar\nu$ SES values.
Note that $n_b$ also includes the $K^+ \to \pi^+ \nu \bar \nu$ events, assuming $\BR_{\nunu}^{\SM}=8.4\times10^{-11}$ as in Ref.~\cite{NA62:2024pjp}\footnote{Alternatively, one may treat $K^+ \to \pi^+ \nu \bar \nu$ events as a separate component $n_\nu$ left as a free parameter. This approach will become viable in experimental analyses as the observed $n_\nu$ increases. For a fixed number of d.o.f., we find that allowing $n_\nu$ to float strengthens the bounds by approximately 5-10\% compared to fixing it to the SM value. However, we adhere to the latter approach to align fully with the procedure outlined in Ref.~\cite{NA62:2020xlg}.}.

Given $n_\text{tot}$, its two components are distributed as a function of the invisible invariant mass squared $m_{\inv}^2$ according to a multinomial likelihood, yielding $\mc L_2$ as
\be
{\cal L}_2=
\prod_{j=1}^{n_\text{obs}} \left[ \frac{n_b}{n_\text{tot}}g_b (m_j^2) + \frac{n_a}{n_\text{tot}}g_a (m_j^2) \right] \ ,
\ee
where $g_{b}(m_\inv^2)$ is a function reproducing the distribution of the background events, and normalized to unit integral over the signal regions. We took this function from published data, i.e. from the right panel of Fig.~10 in Ref.~\cite{NA62:2024pjp}. 
Conversely, $g_a(m_\inv^2)$ is a normal distribution centered at the axion mass value with uncertainty given by the invariant mass resolution at fixed $m_a$, $\sigma_{m_a^2}$, taken from the left panel of Figure 2 of Ref.~\cite{NA62:2020xlg}. Finally, the last likelihood factor, $\mc L_3$, is a Poisson-distributed constraint term for the background yield, written as~\cite{Cousins_2008}
\be
{\cal L}_3=\frac{(\tau n_b)^{n_\text{off}}}{n_\text{off}!}e^{-\tau n_b} \ ,
\ee
with $\tau=\mu_b/\sigma_b^2$ and $n_\text{off}=(\mu_b/\sigma_b)^2$ constructed from the
estimated mean $\mu_b$ and uncertainty $\sigma_b$.
We neglect the uncertainty on the rest of the nuisance parameters considered in Ref.~\cite{NA62:2020xlg}, namely the SES uncertainty, as well as the error on the parameters of the axion mass probability density function. By inspection, these uncertainties are subleading with respect to those discussed, and are expected to affect our results in a minor way.

\begin{figure*}[t]
\begin{center}
\includegraphics[width=0.482\textwidth]{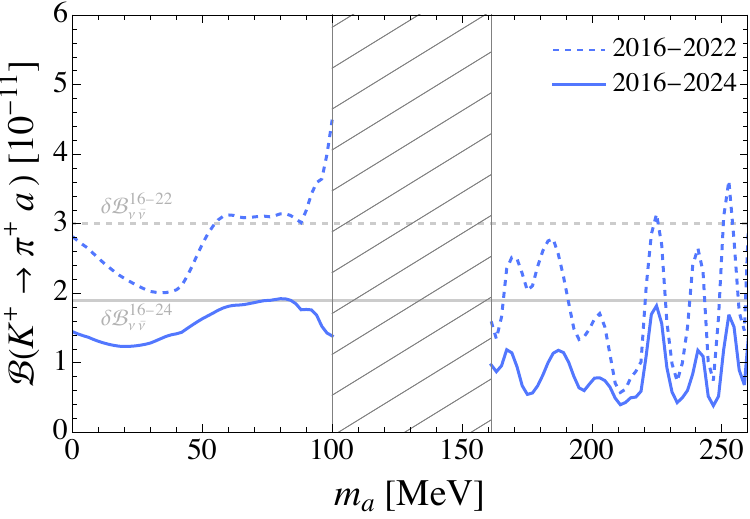} \hfill
\includegraphics[width=0.49\textwidth]{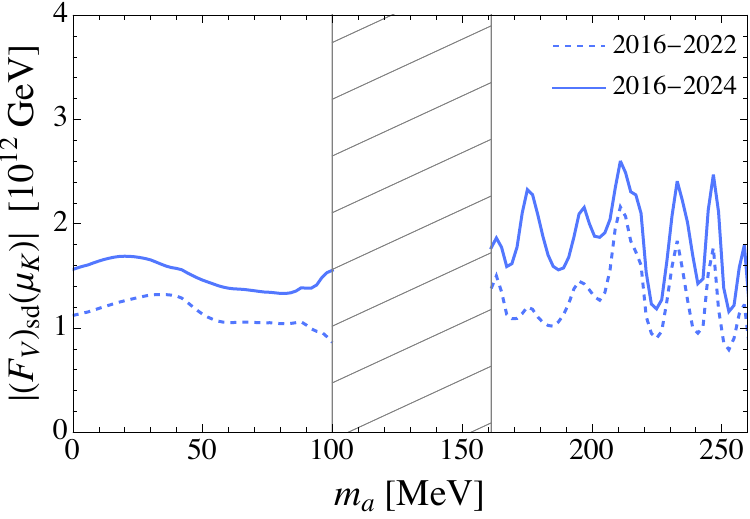}
\caption{(Left) Upper limit on $\BR(K^+\to\pi^+a)$ from NA62 using the published 2016–2022 dataset~\cite{NA62:2025upx}, and the extended 2016–2024 sample including the new 2023–2024 data~\cite{LaThuile_2026_update}. The gray dashed line shows $\delta\BR_{\nunu}$. The range $[100,161]$ MeV is excluded due to $K^+\to\pi^+\pi^0$ background contamination~\cite{NA62:2020xlg,NA62:2024pjp}.
(Right) Corresponding lower bound on the flavour-violating-coupling rescaled Peccei–Quinn scale (see text for details).}
\label{fig:BR_FVsd}
\end{center}
\vspace{-0.5cm}
\end{figure*}
Minimizing $\chi^2(n_b, n_a) \equiv -2\log \mc L$, and demanding $\Delta \chi^2 \equiv \chi^2 - \chi^2_\min = \chi^2(90\%, 2~\dof)$ we obtain a limit on $\BR(\Kppipa)$ as a function of $m_a$. 
We tested this procedure by comparing our obtained limit with the NA62 results based on 2016-2018~\cite{NA62:2021zjw} and 2017 data~\cite{NA62:2020xlg}, respectively, finding very good agreement. The details of this validation, and plots quantifying the agreement, are provided in the supplemental~\refsec{app:crosschecks}. We apply this procedure to two datasets: the full published NA62 dataset, encompassing data from 2016 to 2022, as well as the dataset further including the data recorded in 2023-2024~\cite{LaThuile_2026_update}. We obtain the two BR limits shown in \reffig{fig:BR_FVsd} (left). At $m_a=0$ these limits read
\begin{subequations}
\label{eq:BRexp}
\begin{align}
\BR(K^+\to\pi^+a) &< 2.8 \times 10^{-11}& \hspace{-4ex}\equiv&~\BR_{16\text{--}22} ~, \label{eq:BRexp:a}\\
\BR(K^+\to\pi^+a) &< 1.4 \times 10^{-11}& \hspace{-4ex}\equiv&~\BR_{16\text{--}24} ~,\label{eq:BRexp:b}
\end{align}
\end{subequations}
at 90\% C.L. for the two mentioned datasets.
The bounds \refeq{eq:BRexp} improve over Refs.~\cite{NA62:2021zjw} and~\cite{NA62:2025upx} by a factor of about 2 in both cases, in line with the corresponding increase in the data sample size.

\refcite{NA62:2025upx}, which appeared months after the first version of this study, includes an analysis of $K^+ \to \pi^+ X$ where $X$ is an ALP. At $m_X = 0$, \refcite{NA62:2025upx} agrees with our result in \refeq{eq:BRexp:a} to within one unit of the last significant digit. This agreement constitutes a strong validation of our approach: it confirms the possibility of retrieving the leading uncertainties from public data alone, our treatment of these uncertainties, and the negligible impact of those we did not include. While a full NA62 search remains the benchmark, the above comparison shows that our approach provides an excellent approximation, and has the further key advantage of being fully reproducible. Below we use the state-of-the-art limit \refeq{eq:BRexp:b} to set unprecedented bounds on the relevant axion coupling or, alternatively, on the axion scale.

\noi {\bf Theory Prediction---} %
The axion, as the near-Goldstone boson of a global $U(1)$ Peccei-Quinn symmetry, can be incorporated into ChPT as elucidated in Ref.~\cite{\georgi}. We refer to the $O(p^2)$ ChPT Lagrangian augmented with an axion as the ``strong'' Lagrangian $\Las$. A generalization of this procedure enables inclusion of the effect of strangeness ($S$) changing weak interactions. We denote as $\Law$ the $|\Delta S| = 1$ contributions which are enhanced by the so-called $\Delta I = 1/2$ rule. These include the well-known $G_8$ term {\em and}, in the presence of an axion, the so-called $G_8^\theta$ term~\cite{\cornella}. We find that the coupling $G_8^\theta$ is {\em not} independent of $G_8$. Requiring that the $K^0 \to Z$ amplitude vanishes to the perturbative order $p^2 G_F$ we are considering~\cite{\isidori}, implies the relation $G_8^\theta = - 2 G_8$ up to loop-suppressed penguin contributions. We are thus able to include the $G_8^\theta$ contribution in a fully quantitative way for the first time. The amplitude $\mc M$ for the $K^+ \to \pi^+ a$ decay can be decomposed into two pieces,
\be
\label{eq:M}
\mc M(\boldsymbol{k}) \equiv \Ms((k_V)_{sd}) + \Mw((k_A)_{uu,dd,ss},(k_V)_{dd,ss})~,
\ee
induced by $\Las$ and $\Law$, respectively. We collect explicit formulae for the Lagrangians and for the amplitudes in the Supplemental \refsec{app:ChPT}.

The $k$ couplings in \refeq{eq:M} are understood at a renormalization scale $\mu_K \approx m_K$. A bound from $K \to \pi a$ on these couplings is informative on the strength of the axion-induced flavour-violating mechanism at the energy of the decay, but it says little about the fundamental couplings at the high scale $\muPQ \sim 4 \pi f_a$ where the PQ symmetry is broken.
To this end, one needs to renormalization-group (RG) evolve the $k$ couplings appearing in \refeq{eq:M} to $\muPQ$.

We note that axion-induced flavour violation at low energies can arise not only from the same coupling at high energies, but also from the misalignment between the up- and down-type quark Yukawa couplings $Y_{u,d}$, parameterized by the Cabibbo-Kobayashi-Maskawa matrix $V$. Both these inputs, therefore, have to be carefully included. Using $Y_{u,d}$ and $V$ at the electroweak scale $\mu_{\rm EW} \sim m_t$ (e.g. Ref.~\cite{DiNoi:2022ejg}) as boundary conditions, we obtain their counterparts at the $\muPQ$ scale using \refcite{Arason:1991ic}. We then RG-evolve from $\muPQ$ to $\muEW$ the coupled equations (implemented following the comprehensive \refcite{\neubertRGE}) for Yukawa and axion couplings, using the PQ-scale Yukawa and axion couplings as boundary conditions. The axion couplings amount to 35 degrees of freedom (d.o.f.) in total, and include: the axion-gauge boson couplings $c_{BB}$ and $c_{WW}$~\footnote{For the QCD axion, the coupling to gluons {\em defines} the PQ scale \cite{\diluzio}, see \refeq{eq:Laqq}. For the normalization of the $G \tilde G$ operator, see Supplemental \refsec{app:ChPT}.}; the axion couplings to the chiral matter multiplets $\Phi = Q$, $L$ and $u, d, e$, denoted as $(k_\Phi)_{ij}$, with $i,j$ generation indices---that we distinguish from the light-quark indices used, e.g., around \refeq{eq:Laqq}, by denoting the latter indices with light-quark symbols.

Importantly, the definition of $k$-coupling entries at the PQ scale is quark-field-basis dependent, i.e. flavour-violating effects can be shuffled from non-diagonal Yukawa (i.e. quark-mass) matrices to non-diagonal $k$-coupling matrices. Given that our underlying transition is $s \to d$, the most convenient basis in our case is the down-quark basis, where the down-type Yukawa matrix is diagonal. In this basis PQ-scale dynamics is thus understood to be flavour-conserving or -violating in the down sector~\footnote{In a different basis, off-diagonal entries of the axion-down-type quark coupling matrix are also induced by the misalignment between the gauge and the mass eigenstates, see e.g. the first term on the right side of Eq.~(5.17) of Ref.~\cite{\neubertRGE} for the case of choosing the up-quark basis. More generally, the $k$ couplings are defined up to the bi-unitary transformations of the up- and down-type quark fields that diagonalize the Yukawa matrices. Each of these transformations can induce off-diagonalities even starting from diagonal $k$ matrices.} and $Y_d(\muPQ)$ is, by definition, diagonal.

To the thus-obtained Yukawa and axion couplings at the $\muEW$ scale we add matching contributions at $\muEW$~\cite{\neubertRGE}. Finally, RG effects below $\muEW$ are, depending on the coupling, either absent or tiny, and can be safely neglected~\footnote{Specifically, RG effects in off-diagonal $k$ couplings are suppressed below $\muEW$ by the Fermi constant and the tiny Yukawa couplings of light quarks~\cite{\neubertRGE,\neubertALPslong}; the same effects in diagonal $k$ couplings enter $\Mw$ only, and thus amount to contributions to the amplitude of $O(G_F^2)$.}.

\noi {\bf Results---}%
We next discuss the implications of the relation $\mc B(|\mc M|) \le \mc B_{\exp}$, with $\mc M$ from \refeq{eq:M}. Here $\Ms$ is proportional to the relevant flavour-violating axion coupling $(k_V)_{sd} (\muK)$. Conversely, $\Mw$ depends on flavour-conserving $k$ couplings only, and has an overall parametric suppression of order $G_F F_0^2 \approx 1.0 \times 10^{-7}$.
Therefore, the natural expectation is that strong contributions to $K \to \pi a$ are larger than weak contributions, i.e. $|\Ms| \gg |\Mw|$, in which case the experimental bounds \refeq{eq:BRexp} translate into bounds on $|(k_V)_{sd} (\muK)| / (2f_a)$ or, equivalently, its inverse, that for $m_a = 0$ read
\begin{subequations}
\label{eq:FVsdbound}
\begin{align}
\label{eq:FVsdbound:a}
|\FV{sd} (\mu_K)| > 1.1 \times 10^{12}~\GeV~,\\
\label{eq:FVsdbound:b}
|\FV{sd} (\mu_K)| > 1.6 \times 10^{12}~\GeV~,
\end{align}
\end{subequations}
These two bounds are 1.6 and 2.4 times, respectively, stronger than 
the latest data-driven bound~\cite{\camalich} on the effective scale of the underlying interaction. In \reffig{fig:BR_FVsd} (right) we show a generalization of \refeq{eq:FVsdbound} for any $m_a$ in the range of values probed by NA62.

Our \reffig{fig:BR_FVsd} assumes a $\pi^+$ plus missing-energy final state, i.e. that the axion escapes the approximately $130$~m-long instrumented region downstream of the kaon decay without decaying visibly. In the mass range shown in the figure, the axion could in principle decay into $\gamma\gamma$, $e^+e^-$, or $\mu^+\mu^-$ (decays into hadrons require the two-pion threshold). For $m_a \lesssim 1~\MeV$, only the $a\to\gamma\gamma$ channel is kinematically allowed. A sufficient condition for the axion to escape detection within the instrumented region is $|g_{a\gamma\gamma}| \lesssim 10^{-8}$, which ensures a decay length larger than the detector size. Existing bounds on $g_{a\gamma\gamma}$~\cite{AxionLimits} are significantly stronger than this value, so our assumption is safely satisfied in the regime $m_a \lesssim 1\,\MeV$. For $m_a \gtrsim 1\,\MeV$, the $m_a$–$f_a$ relation is strongly model-dependent. Rather than entering model-specific considerations, we provide the data-driven requirement $\tau_a \gtrsim 10$~ns~\cite{NA62:2025upx}, which can be imposed on any given model.

In QCD axion models $m_a$ is uniquely determined by $f_a$. Expressing $f_a$ through $(F_V)_{sd}$ and $(k_V)_{sd}$, and inserting the bound \refeq{eq:FVsdbound}, one finds that for a natural coupling $|(k_V)_{sd}|\sim 1$ the corresponding axion mass lies in the range $m_a \sim 10^{-15}$--$10^{-14}\,\GeV$. Reducing $|(k_V)_{sd}|$ to $O(10^{-7})$ (i.e.\ $O(G_F F_0^2)$) raises the implied mass to $m_a \sim 10^{-7}\,\GeV$ (see also suppl. \refsec{app:ma_vs_ksd}). For even smaller values of $|(k_V)_{sd}|$, the strong and weak contributions to the amplitude become comparable and one enters the finely tuned regime discussed below. It follows that over the relevant parameter range the axion is effectively massless on accelerator scales: both the dynamical scale of the process ($m_K \sim 500\,\MeV$) and the experimental resolution exceed the implied QCD-axion mass by many orders of magnitude. Although accelerator limits are therefore insensitive to such small $m_a$ values, they impose strong constraints on the flavour-violating coupling and thereby indirectly single out a well-defined QCD-axion mass window.

Based on the discussion after \refeq{eq:M}, we can also study the inequality $\mc B(|\mc M|) \le \mc B_{\exp}$ directly in terms of the UV couplings at the scale $\mu_\PQ$. Since the amplitude is linear in $1/f_a$, this inequality can be rewritten as:
\be
\label{eq:fa}
f_a \ge \sqrt{\frac{\mc B(f_a \cdot |\mc M|)}{\mc B_{\exp}}}~,
\ee
where the r.h.s. depends on the UV-scale $k$ couplings only---i.e. it does {\em not} depend on $f_a$. This leads to
\be
\label{eq:faz}
f_a \ge \underbrace{4.85 \times 10^4~\GeV}_{\mbox{defines~} \fab} \cdot \underbrace{|1 + \boldsymbol{\alpha}\cdot\mathbf{c}|}_{\mbox{defines~} |z|}~,
\ee
where $\mathbf{c}$ is the vector of 35 couplings at $\mu_\PQ$ and $\boldsymbol{\alpha}$ is the vector of multiplying coefficients. The unit term inside the absolute-value factor in \refeq{eq:faz} tracks the contribution to $\mc B_a$ due to the $a G \tilde G$ term in \refeq{eq:Laqq}: the $c_i \to 0$ limit yields the bound $f_a \ge \fab$.

The factor $\az$ in \refeq{eq:faz} reflects the coexistence of strong and weak contributions, the latter being parametrically suppressed by $O(10^{-7})$. Accordingly, the coefficients $\alpha_i$ span several orders of magnitude, from $O(10^{-2})$ up to $O(10^{6})$ (a complete tabulation at the reference scale $\mu_\PQ = 10^{12}~\GeV$ is provided in Table~\ref{tab:aici} of the supplemental \refsec{app:fa:alphas}).
No established theory fixes the UV values of the couplings $c_i$, but they plausibly lie in the range $[-5,5]$. (Larger $|c_i|$ would raise perturbativity concerns; moreover, the largest SM Yukawa coupling---conceptually similar to the couplings in \refeq{eq:Laqq}—is $\sim 1$.)
For generic $c_i$ values in this range one finds $|z|\sim \az_{\rm peak} \equiv 1 \cdot 10^{8}$, implying typical bounds $f_a^{\rm gen} \sim 10^{12}~\GeV$. This becomes $f_a^{\rm FC} \sim 10^{8}~\GeV$ if one assumes flavour-violating couplings to be zero at the PQ scale, representing theories where the axion is flavour-conserving at that scale. In both cases, illustrated in the supplemental \reffig{fig:fa_distr}, the $f_a$ bound is thus orders of magnitude stronger than $\bar f_a$.

Achieving $\az \sim O(1)$ (weak-amplitude dominance) requires cancellations among many complex contributions: geometrically, it corresponds to forcing a complex quantity whose typical magnitude is $\sim 10^{7}$ to lie inside a disk of radius $O(1)$ around the origin. For random UV choices $c_i \in [-5,5]$, this region has an exceedingly small measure. For $r \ll \az_{\rm peak}$ the small-$\az$ probability scales as $r^2$, which allows an extrapolation of the Monte Carlo results described in the Supplemental~\refsec{app:fa:scaling}. We obtain $P(\az \le 1) \sim 10^{-16}$, implying that $O(10^{16})$ trials are required for a single instance of $\az\lesssim 1$.

We therefore infer the bound
\be
\label{eq:fa_bound}
f_a \ge \fab = 4.9 \times 10^4~\GeV~,
\ee
which should be regarded as a general and conservative lower limit: in the vast majority of parameter space the implied bound on $f_a$ is much stronger. The limit in \refeq{eq:fa_bound} corresponds to $m_a \le 1.3 \times 10^{-7}\,\GeV$.

Applying to the weak-amplitude-dominated case, the bound in \refeq{eq:fa_bound} is sensitive to the choice of weak couplings $g_8$ and $G_8^\theta$---see Suppl.~\refsec{app:ChPT} for their definition and the reference values used in this work. For example, using the ``classic'' $g_8 \simeq 5$ or instead 3.61~\cite{Cirigliano:2011ny}, and in the absence of $G_8^\theta$~\cite{\cornella}, whose effect we are in a position to quantitatively include for the first time in this work, one would obtain $1.70\times 10^5$ or $1.35\times 10^5\,\GeV$, respectively.

While $f_a\ge\bar f_a$ is a conservative lower limit, we also examined explicitly two specific corners of parameter space in which the bound in \refeq{eq:fa_bound} can be approached: single-coupling tuning, and engineered cancellations between one parameter and the sum of the rest. A detailed analysis is presented in the Supplemental \refsec{app:fa:cases_i_and_ii}. We find that these configurations occupy a negligibly small region of the allowed UV parameter space and, moreover, require relations connecting quantities that are structurally unrelated, such as UV couplings and coefficients generated over many decades of RG evolution.

\noi {\bf Outlook---}%
A salient feature of flavoured accelerator observables is the coexistence of strong and weak contributions to the flavour-changing amplitude, with the weak piece parametrically suppressed. This implies two conceptually distinct constraints on the PQ scale, the strongest constraint arising in the generic regime of strong-amplitude dominance. Conversely, weak dominance implies a weaker limit but requires highly tuned UV coupling configurations. Precisely the difficulty of realizing such configurations renders the corresponding bound valuable as a conservative general lower limit on $f_a$. Moreover, these constraints are obtained in a fully controlled laboratory environment and thus complementary to astrophysical constraints.

The analysis described above can be extended to other kaon decays, notably modes with a pion pair plus missing energy, which probe the complementary $(k_A)_{sd}$ coupling. In the $K^+ \to \pi^+ \pi^0 a$ channel, NA62 is expected to reach $\mc B \sim 10^{-7}$~\cite{Goudzovski:2022vbt}, two orders of magnitude below the current best limit~\cite{Tchikilev:2003ai}. Relative to the one-pion-plus-invisible case (\reffig{fig:BR_FVsd}), the additional $\pi^0$ reduces sensitivity through the smaller trigger bandwidth (by $\sim1/400$), the geometric acceptance of the two photons in the ECAL ($\sim1/4$), and the larger background in three- versus two-body kinematics ($\sim1/10$), while reconstruction efficiencies for $\pi^+$ and $\pi^0$ remain comparable. Altogether this implies $\mc B(K^+ \to \pi^+ \pi^0 a)\lesssim 5\times10^{-7}$. For the analogous neutral channel, E391a has set $\mc B(K_L \to \pi^0 \pi^0 a) < 7 \times 10^{-7}$ at 90\% CL~\cite{E391a:2011aa}, and future improvements are expected from KOTO and KOTO-II~\cite{KOTO:2025gvq}. A sensitivity at the level of $\mc B(K\to\pi\pi a) \lesssim 5 \times 10^{-7}$ therefore provides a realistic benchmark, translating into
\be
|\FA{sd}| > 1.0 \times 10^{8}~\GeV~,
\label{eq:FAsdbound_indicative}
\ee
which exceeds the latest bound in Ref.~\cite{\camalich} by one order of magnitude. As for the vector case, this constraint is noteworthy both for its strength and for being derived in a controlled experimental setting.

\noi {\bf Acknowledgments---} 
We acknowledge discussions with Maël Cavan-Piton in the initial stages of the project. A special thanks goes to Claudia Cornella for key clarifications and insights on $G_8^\theta$. This work has received funding from the French ANR, under contracts ANR-19-CE31-0016 (`GammaRare') and ANR-23-CE31-0018 (`InvISYble'), that we gratefully acknowledge.

\bibliography{bibliography}

\onecolumngrid

\newpage

\begin{center}
\large \bf Supplemental Material
\end{center}

\appendix
\renewcommand\appendixname{}

\renewcommand{\thesubsection}{\thesection\arabic{subsection}}

\section{Validations of our reinterpretation procedure} \label{app:crosschecks}

\noi We tested the procedure described in the main text by applying it to 2017~\cite{NA62:2020xlg} and 2016-2018 data~\cite{NA62:2021zjw}, and verifying that the obtained upper limits on $\BR(K^+\to\pi^+a)$ reproduce the published counterparts in Refs. \cite{NA62:2020xlg,NA62:2021zjw}.

\begin{figure}[h!]
\begin{center}
\includegraphics[scale=0.65]{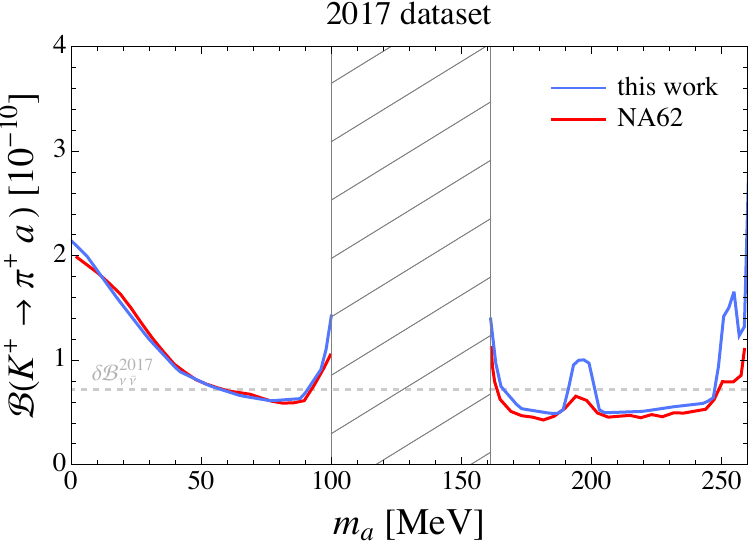}
\hspace{8mm}
\includegraphics[scale=0.65]{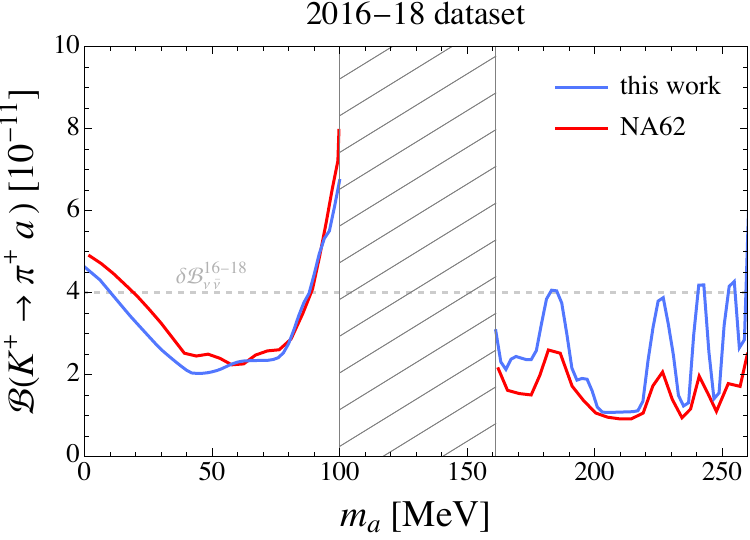}
\caption{Upper limit on $\BR(K^+\to\pi^+a)$ from the 2017 \cite{NA62:2020xlg} and 2016-18 \cite{NA62:2021zjw} datasets. The solid red line is the published result of NA62 collaboration. The gray dashed line shows the corresponding value of $\delta\BR_{\nunu}$ for comparison.}
\label{fig:crosscheck2}
\end{center}
\end{figure}

In Fig.~\ref{fig:crosscheck2} we show our results obtained for the accessible values of axion masses compared to the published ones from the collaboration. We find a very good agreement, to within a few percent, if we compare to the $\chi^2$ value with 90\% probability and 2 dof, which we understand to be due to neglecting the uncertainty of the knowledge of the SES.

\section{\boldmath ChPT conventions and $K^+ \to \pi^+ a$ amplitudes} \label{app:ChPT}

\noi We collect the Lagrangians relevant to our calculations. The ``strong'' Lagrangian reads
\begin{align}
\label{Lstrong}
&\Las = \frac{F_0^2}{4} \Bigl( \Tr D_\mu U (D^\mu U)^\dagger + 2 B_0 \Tr(\hat M_q U + U^\dagger \hat M_q^\dagger ) \Bigl)~.
\end{align}
The pion field $U \equiv \exp(i \frac{\phi^a}{F_0} \lambda^a)$ is normalized such that $(\phi^a \lambda^a)_{11} = \pi^0 + \eta_8 / \sqrt3$, with $F_0 = 93\,\MeV$, and transforms as $U \rightarrow R U L^\dagger$; 
$D_\mu U = \partial_\mu U - i r_\mu U + i U l_\mu$, with $r_\mu$, $l_\mu$ external currents, e.g.
\begin{align}
&r_\mu^{\rm e.m.} = l_\mu^{\rm e.m.} = - e A_\mu Q~, \nonumber \\
&l_\mu^{W} = - \frac{g}{\sqrt2}\, W_\mu^+ \begin{pmatrix}
0 & V_{ud} & V_{us} \\
0 & 0 & 0 \\
0 & 0 & 0
\end{pmatrix} + \hc~,~~ r_\mu^{W} = 0~,\nonumber\\
&r_\mu^{Z} = - \frac{g}{\cos\theta_w}\, Z_\mu^0 (- Q \sin^2 \theta_w)~, ~~ 
l_\mu^{Z} = - \frac{g}{\cos\theta_w}\, Z_\mu^0 (T_3 - Q \sin^2 \theta_w)~.
\end{align}
with $Q = \diag(Q_u, Q_d, Q_s)$ and $T_3 = \diag(1/2, -1/2, -1/2)$.

As elucidated in Ref.~\cite{\georgi}, the above Lagrangian assumes that the $a G \tilde G$ coupling in \refeq{eq:Laqq} has been removed in favour of an axion-dependent phase in the quark mass matrix $M_q = \diag(m_u, m_d, m_s)$. We define this quark redefinition as $q \to \exp[i (\delta +\kappa \gamma_5) a / (2 f_a)] q$, with $\Tr \kappa = 1$ and the hermitian matrices $\delta, \kappa$ chosen as diagonal, implying \cite{\georgi,\neubertPRL}
\be
\hat M_q = \exp\left(i \frac{a}{f_a} \kappa \right) M_q~, \quad
r_\mu^{a} = \frac{\partial_\mu a}{f_a}\, \hat k_{R}(a)~, \quad l_\mu^{a} = \frac{\partial_\mu a}{f_a}\, \hat k_{L}(a)~,
\ee
with $\hat k_{R,L}(a) = e^{- i a / (2f_a) (\delta \pm \kappa)} \bigl(k_{R,L} - (\delta \pm \kappa)/2 \bigl) e^{i a / (2f_a) (\delta \pm \kappa)}$.
Note that the $\kappa$ and $\delta$ dependence cancels out in physical amplitudes such as \refeqs{eq:Ms}{eq:Mw}, as expected. For $G \tilde G$ we use the normalization $\tilde G_a^{\mu \nu} \equiv \frac{1}{2}\epsilon^{\mu \nu \rho \sigma} G_{\rho \sigma, a}$ with $\epsilon^{0123} = +1$ as in Ref.~\cite{Peskin:1995ev}. Our effective scale $f_a$ definition thus coincides with the one in Refs.~\cite{\georgi,DiLuzio:2020wdo}.

The $|\Delta S| = 1$ Lagrangian is given by~\cite{\cornella}
\begin{align}
\label{LDeltaS1}
&\Law = 4 G_8 \left( L_\mu L^\mu \right)^{32} + 2 F_0^2 G_8^\theta
(D_\mu \theta)
(L^\mu)^{32} + \hc~,
\end{align}
with $G_8^{(\theta)} \simeq - \frac{G_F}{\sqrt2} V^*_{ud} V_{us} \, g_8^{(\theta)}$ and with the $L_\mu$ current
\be
L_\mu^{ji} = i \frac{F_0^2}{2} \exp{- i a / (2 f_a) \,\left( (\delta_i - \delta_j) - (\kappa_i - \kappa_j) \right)} \left[ (D_\mu U)^\dagger U \right]^{ji}~.
\ee
The covariant derivative of the external current $\theta$ has been introduced as in Ref.~\cite{\cornella} and is defined as
\be
D_\mu \theta\equiv \partial_\mu \theta - \< r_\mu - l_\mu \> = - \< r_\mu - l_\mu \>~,
\ee
where the last equality comes from the fact that the axion-dependent quark redefinition introduced above set $\theta=0$.
As discussed in the main text, we find that the couplings in the two operators in \refeq{LDeltaS1} are not independent, as they fulfill the relation $G_8^\theta = -2 G_8$, or equivalently $g_8^\theta = -2 g_8$.

Throughout, we mostly follow the conventions in Refs.~\cite{\neubertPRL,\neubertALPslong}, with the exception of the convention on the chiral transformation properties of the meson field, the pion decay constant, and the $g_8$ definition, for which we comply with classical ChPT papers such as Ref.~\cite{\ecker}. Numerically, we take $g_8 = 3.61$~\cite{Cirigliano:2011ny}.

\bigskip

For the $K^+ \to \pi^+ a$ amplitudes we find
\begin{align}
\label{eq:Ms}
i \Ms = 
& \frac{m_K^2 (k_V)_{sd}}{2 f_a}\Bigl( 1 - x_{\pi K}\Bigr)~, \\[3ex]
\label{eq:Mw}
i \Mw = 
& -\frac{N_8^*}{4 f_a} 
\left(
\frac{m_K^4 \, C_{K} + m_K^2 m_\pi^2 \, C_{K\pi} + m_\pi^4 \, C_{\pi} + m_a^2 m_K^2 \, C_{aK} + m_a^2 m_\pi^2 \, C_{a\pi} + m_a^4 \, C_a}{4 m_K^2 - m_\pi^2 - 3 m_a^2}
\right) \nn \\
& \hspace{0.05cm} - \frac{N_8^{\theta*}}{2 f_a}(m_K^2 - m_\pi^2)\left(-1+(k_A)_{uu} + (k_A)_{dd} + (k_A)_{ss}\right) \nn \\
= & - \frac{N_8^{*} m_K^2}{4 f_a}
\Bigl(
-2 + 2 (k_{A})_{uu} + (k_{A})_{dd} + (k_{A})_{ss} + (k_{V})_{dd} - (k_{V})_{ss}
+ 2 \frac{N_8^{\theta\,*}}{N_8^{*}}
\left(-1 + (k_{A})_{uu} + (k_{A})_{dd} + (k_{A})_{ss}\right) \nn \\ 
& \hspace{1.7cm}+ O\left(x_{aK},x_{\pi K}\right)
\Bigr)~,
\end{align}
with $x_{ij} \equiv (m_i / m_j)^2$ and $N_8^{(\theta)} \equiv 2 F_0^2 G_8^{(\theta)}$. In the first line for $i \Mw$, contributions are ordered by coupling ($N_8$ vs $N_8^\theta$) and by expected numerical significance in the QCD-axion case ($x_{a\pi} \ll 1$), and bearing in mind that $x_{\pi K} \ll 1$. The last line is an explicit expansion in these small ratios. The $C$ symbols encode the dependence on the fundamental axion-quark couplings as
\begin{align}
&C_{K} = 4 \Bigl(-2 + 2 (k_A)_{uu} + (k_A)_{dd} + (k_A)_{ss} + (k_V)_{dd} - (k_V)_{ss} \Bigl)~, \nn \\[1ex]
&C_{K\pi} = 8 -10 (k_A)_{uu} - 5 (k_A)_{dd} - 5 (k_A)_{ss} + 3 (k_V)_{dd} - 3 (k_V)_{ss}~, \nn \\[1ex]
&C_{\pi} = 2 (k_A)_{uu} + (k_A)_{dd} + (k_A)_{ss} - (k_V)_{dd} + (k_V)_{ss}~, \nn \\
&C_{aK} = 8 -8 (k_A)_{uu} - (k_A)_{dd} - 3 (k_A)_{ss} - 7 (k_V)_{dd} + 7 (k_V)_{ss}~, \nn \\
&C_{a\pi} = 2 \Bigl(-4 + 4 (k_A)_{uu} + 2 (k_A)_{dd} - (k_V)_{dd} + (k_V)_{ss}\Bigl)~, \nn \\
&C_{a} = -3 \Bigl((k_A)_{dd} - (k_A)_{ss} - (k_V)_{dd} + (k_V)_{ss}\Bigl)~.
\end{align}
In the above equations, the $k$ couplings are understood to be at a renormalization scale $\mu$ close to the scale of the process, i.e. at $\mu_K \approx m_K$.

\section{\boldmath Mass–coupling relation} \label{app:ma_vs_ksd}

\begin{figure}[t]
\begin{center}
\includegraphics[width=0.45\textwidth]{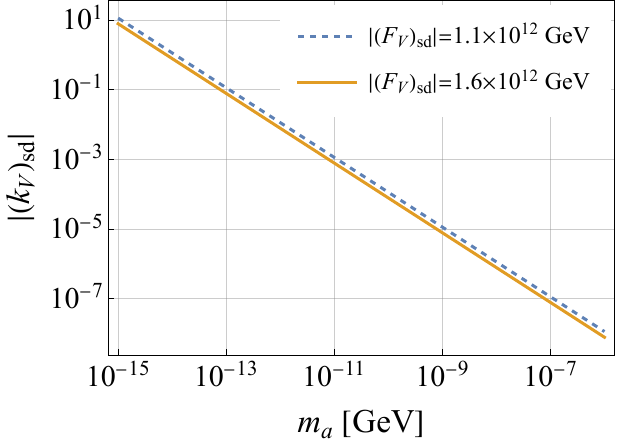}
\caption{QCD axion mass $m_a$ as a function of the effective flavor-violating coupling $(k_V)_{sd}$.}
\label{fig:kV_ma}
\end{center}
\end{figure}
\noi In QCD axion models the mass can be related to the effective flavour-violating scale bounded in this work through
\be
\label{eq:ma_FVkV}
m_a = \frac{\sqrt2 F_\pi m_\pi}{|(F_V)_{sd}| |(k_V)_{sd}|}
\frac{\sqrt{m_u m_d}}{m_u + m_d}~,
\ee
with $F_\pi = 93 \sqrt2~\MeV$. This relation allows to display the impact of the $(F_V)_{sd}$ bound \refeq{eq:FVsdbound} in the mass–vs-coupling plane. This is shown in \reffig{fig:kV_ma}, where we restrict to $|(k_V)_{sd}|\lesssim 10$ to remain within the perturbative regime.

\section{\boldmath $f_a$ limit as a function of the UV couplings} \label{app:fa}

\subsection{Linear form and hierarchy of coefficients}
\label{app:fa:alphas}

\noi The bound on the axion decay constant derived in the main text can be written in the compact form
\be
\label{eq:z_def_app}
f_a \ge \fab \, |z|~,
\qquad
z \equiv 1 + \boldsymbol{\alpha}\cdot\mathbf{c}~,
\ee
where $\mathbf{c}\equiv (c_1,\dots,c_{35})$ denotes the set of 35 dimensionless axion couplings defined at the PQ scale $\mu_{\PQ}$, and the numerical coefficients $\alpha_i$ subsume both strong and weak contributions to the $K\to\pi a$ amplitude, and include  renormalization-group evolution to the scale $\mu_K$ of the decay.

The full list of coefficients $\alpha_i$, evaluated at the reference scale $\mu_{\PQ}=10^{12}~\GeV$, is reported in Table~\ref{tab:aici}. A salient feature is the very wide hierarchy in their magnitudes: $|\alpha_i| \in \left[ O(10^{-2}),\, O(10^{6}) \right]$.
This hierarchy directly reflects the coexistence of two qualitatively different contributions to the decay amplitude: a strong contribution, proportional to the flavour-violating axion coupling $(k_V)_{sd}$, and a weak contribution, which comes with a relative parametric suppression of order $G_F F_0^2 \sim 10^{-7}$. The largest values of $|\alpha_i|$ are associated with couplings that feed the strong amplitude through renormalization-group mixing, while the smallest coefficients correspond to purely weak or electroweak-induced effects.

\refeq{eq:z_def_app} makes explicit that the bound on $f_a$ depends linearly on the UV couplings through the complex quantity $z$. In the absence of special correlations among the components of $\mathbf{c}$, the magnitude of $z$ is therefore expected to be set by the largest coefficients in \reftab{tab:aici}. The implications of this structure for the typical size of $|z|$, and for the plausibility of accidental cancellations leading to $|z|\sim O(1)$, are analyzed in detail in the following subsections.
\begin{table}[h]
\centering
\caption{
Coefficients $\alpha_i$ of the linear form $z = 1 + \boldsymbol{\alpha}\cdot\mathbf{c}$ at the reference scale $\mu_\PQ = 10^{12}~\GeV$, ordered by decreasing magnitude $|\alpha_i|$. Phases are given in radians.
}
\label{tab:aici}
\renewcommand{\arraystretch}{1.12}
\footnotesize

\begin{tabular}{cccc}

\begin{minipage}[t]{0.24\textwidth}\centering
\begin{tabular}{|c|c|c|}
\hline
$c_i$ & $|\alpha_i|$ & $\arg(\alpha_i)$ \\
\hline\hline
$c_{d12}^{\mathrm{im}}$ & $4.65\times10^{6}$ & $+1.678$ \\
$c_{d12}^{\mathrm{re}}$ & $4.65\times10^{6}$ & $-1.463$ \\
$c_{Q12}^{\mathrm{im}}$ & $4.65\times10^{6}$ & $+1.678$ \\
$c_{Q12}^{\mathrm{re}}$ & $4.65\times10^{6}$ & $-1.463$ \\
\hline
$c_{Q13}^{\mathrm{im}}$ & $2.78\times10^{4}$ & $+1.482$ \\
$c_{Q13}^{\mathrm{re}}$ & $2.78\times10^{4}$ & $-1.659$ \\
\hline
$c_{Q23}^{\mathrm{re}}$ & $5.86\times10^{3}$ & $+0.281$ \\
$c_{Q23}^{\mathrm{im}}$ & $5.86\times10^{3}$ & $-1.292$ \\
\hline
\end{tabular}
\end{minipage}

&

\begin{minipage}[t]{0.24\textwidth}\centering
\begin{tabular}{|c|c|c|}
\hline
$c_i$ & $|\alpha_i|$ & $\arg(\alpha_i)$ \\
\hline\hline
$c_{d13}^{\mathrm{re}}$ & $8.27\times10^{2}$ & $-0.089$ \\
$c_{d13}^{\mathrm{im}}$ & $8.27\times10^{2}$ & $+1.482$ \\
\hline
$c_{Q22}$ & $2.70\times10^{2}$ & $-2.848$ \\
$c_{Q11}$ & $1.16\times10^{2}$ & $+0.293$ \\
$c_{u33}$ & $8.56\times10^{1}$ & $+0.300$ \\
$c_{Q33}$ & $7.00\times10^{1}$ & $+0.300$ \\
\hline
$c_{d22}$ & $2.02\times10^{1}$ & $-2.842$ \\
$c_{d11}$ & $1.90\times10^{1}$ & $+0.311$ \\
\hline
$c_{d23}^{\mathrm{re}}$ & $9.06$ & $+0.281$ \\
$c_{d23}^{\mathrm{im}}$ & $9.06$ & $-1.292$ \\
\hline
\end{tabular}
\end{minipage}

&

\begin{minipage}[t]{0.24\textwidth}\centering
\begin{tabular}{|c|c|c|}
\hline
$c_i$ & $|\alpha_i|$ & $\arg(\alpha_i)$ \\
\hline\hline
$c_{u23}^{\mathrm{re}}$ & $1.98\times10^{1}$ & $-3.055$ \\
$c_{u23}^{\mathrm{im}}$ & $3.81$ & $-2.807$ \\
\hline
$c_{u22}$ & $0.733$ & $-0.260$ \\
$c_{u11}$ & $0.995$ & $-3.137$ \\
$c_{d33}$ & $0.269$ & $-2.849$ \\
$c_{WW}$ & $0.104$ & $-2.843$ \\
\hline
$c_{u13}^{\mathrm{im}}$ & $0.0893$ & $+1.499$ \\
$c_{u13}^{\mathrm{re}}$ & $0.0815$ & $-0.107$ \\
\hline
\end{tabular}
\end{minipage}

&

\begin{minipage}[t]{0.24\textwidth}\centering
\begin{tabular}{|c|c|c|}
\hline
$c_i$ & $|\alpha_i|$ & $\arg(\alpha_i)$ \\
\hline\hline
$c_{L11}$ & $0.0331$ & $+0.296$ \\
$c_{L22}$ & $0.0331$ & $+0.296$ \\
$c_{L33}$ & $0.0311$ & $+0.296$ \\
\hline
$c_{BB}$  & $7.73\times10^{-3}$ & $-2.845$ \\
$c_{e11}$ & $7.73\times10^{-3}$ & $-2.845$ \\
$c_{e22}$ & $7.72\times10^{-3}$ & $-2.846$ \\
$c_{e33}$ & $5.79\times10^{-3}$ & $-2.844$ \\
\hline
$c_{u12}^{\mathrm{im}}$ & $8.42\times10^{-3}$ & $+1.678$ \\
$c_{u12}^{\mathrm{re}}$ & $7.57\times10^{-3}$ & $-3.035$ \\
\hline
\end{tabular}
\end{minipage}

\end{tabular}
\end{table}

\subsection{Generic behaviour and geometric interpretation}
\label{app:fa:scaling}

\noi We analyze here the generic behaviour of $z$, as defined in \refeq{eq:z_def_app}, entering the bound \refeq{eq:faz}, under minimal assumptions on the UV couplings $\mathbf{c}$.

\medskip

\begin{figure}[t!]
\begin{center}
\includegraphics[width=0.49\textwidth]{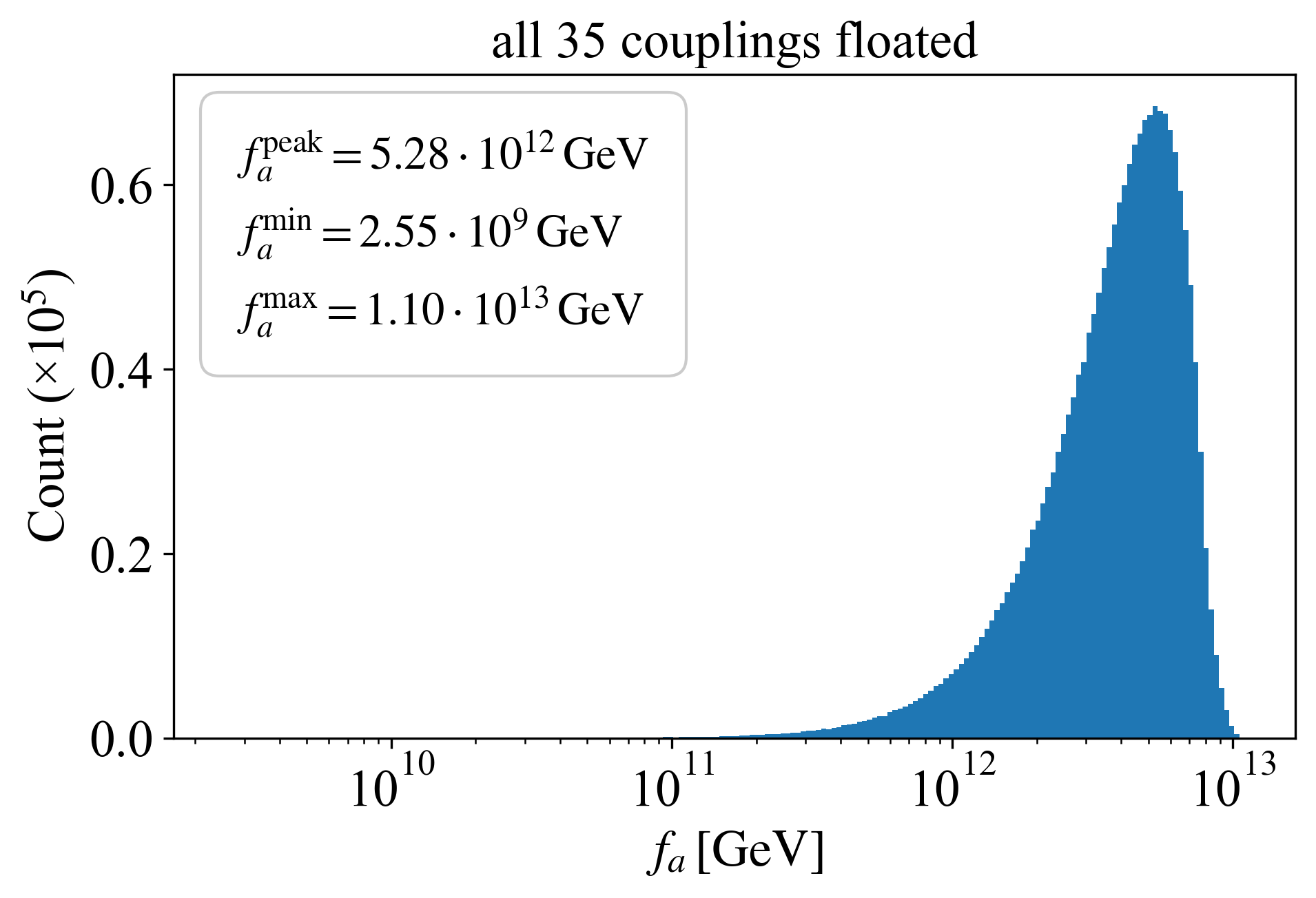}
\includegraphics[width=0.49\textwidth]{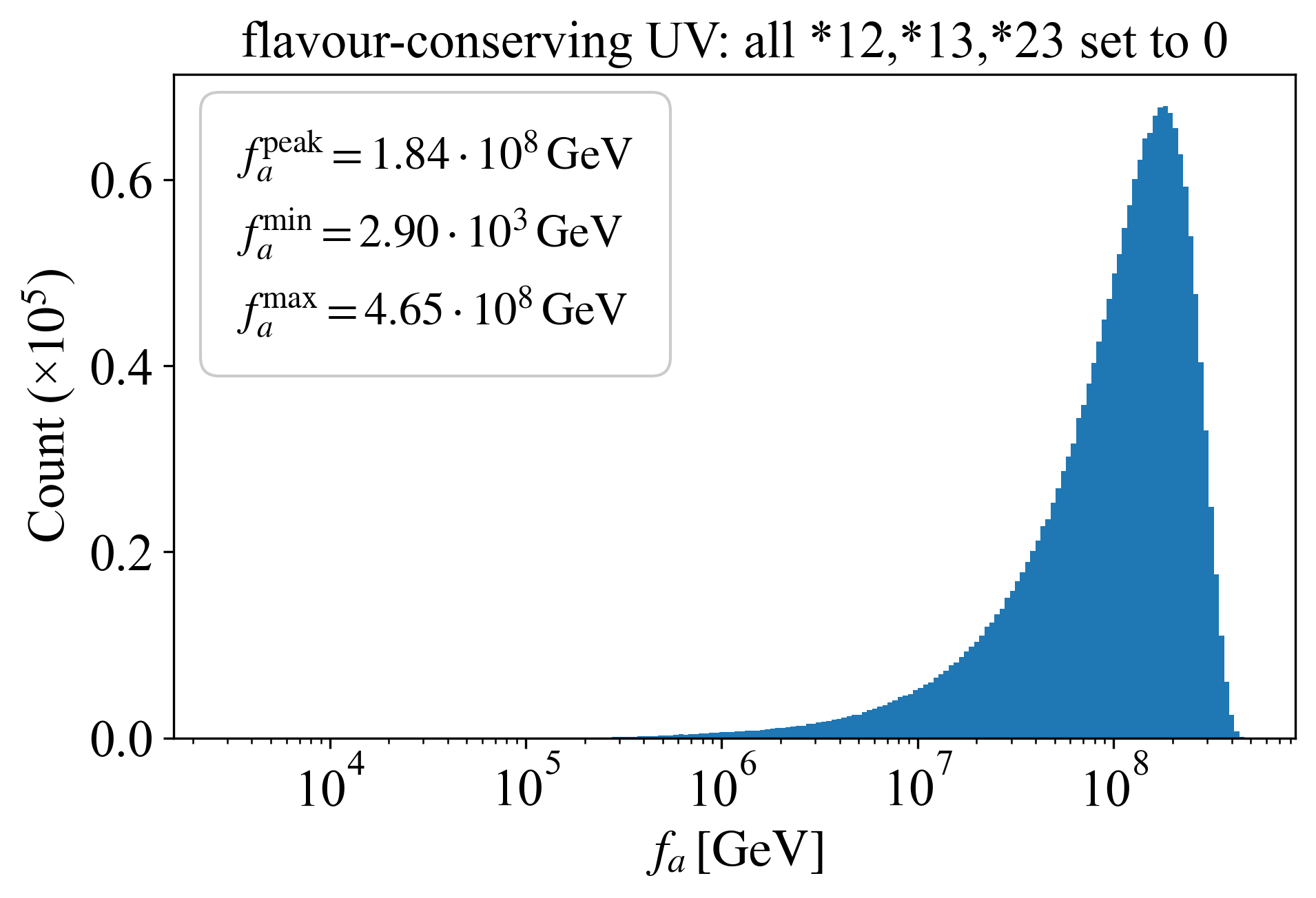}
\caption{Distribution of $f_a$ bounds as defined in \refeq{eq:faz}, assuming $c_i \sim \Unif[-5,5]$: (left) scan of all UV couplings; (right) scan with UV flavour-violating couplings set to zero.}
\label{fig:fa_distr}
\end{center}
\end{figure}

\noi {\em Typical scale of the amplitude.} As shown in \reftab{tab:aici}, the coefficients $\alpha_i$ span several orders of magnitude, up to $|\alpha_i|\sim O(10^{6})$. For bounded UV couplings $c_i\in[-5,5]$ with no special correlations, the linear combination $\adotc$ therefore has a typical magnitude $\azexpl \sim O(10^{7})$, so that $\az$ is generically dominated by this term rather than by the constant~1.
This expectation is confirmed by the numerical distribution shown in \reffig{fig:fa_distr} (left).

\medskip

\noi {\em Induced distribution in the complex plane.} Since $z$ is a linear function of the real variables $c_i$, any choice of a joint distribution for the $c_i$ with a smooth density and finite variance induces a smooth probability density $f(z)$ for the complex variable $z$. No assumption of Gaussianity or independence is required for what follows; only local smoothness of $f(z)$ near the origin will be used.

\medskip

\noi {\em Small--$|z|$ scaling.} The ``small-ball'' probability for $z$ to lie inside a disk of radius $r$ around the origin is
\be
P(\az \le r) = \int_{\az \le r} f(z)\, d^2 z = \pi r^2 f(0) + o(r^2) \qquad (r\to 0)~,
\ee
where the first equality is by definition and the second follows from a Taylor expansion, assuming only smoothness of $f(z)$ at the origin and otherwise independently of its detailed form. For $r \ll \az_{\rm peak}$ the cumulative probability therefore scales as $r^2$. This behaviour is purely geometric, reflecting the area of a disk in the complex plane and not relying on any further distributional assumption.

\medskip

\noi {\em Extrapolation from finite scans.} The quadratic scaling implies that the probability of reaching the weak-amplitude regime $\az \sim O(1)$ is suppressed by the square of the ratio between the target radius---unity---and the typical radius---$\az _{\rm peak} = O(10^7)$---of the distribution. This can be estimated {\em empirically} by determining $P(\az \le r_0)$ for intermediate values $1 \ll r_0 \ll \az_{\rm peak}$ and extrapolating according to $P(\az \le r)\propto r^2$.
Applying this procedure to our numerical scan shows the onset of a stable plateau for $r_0\lesssim 10^{6}$, corresponding to
\be
P(\az \le 1) ~\simeq~ 1\times 10^{-16}~,
\ee
as also illustrated quantitatively by \reffig{fig:fa_distr} (right), and by the accompanying \reftab{tab:small_ball}.
\begin{figure}[t!]
\vspace{-0.5cm}
\begin{center}
\includegraphics[width=0.53\textwidth]{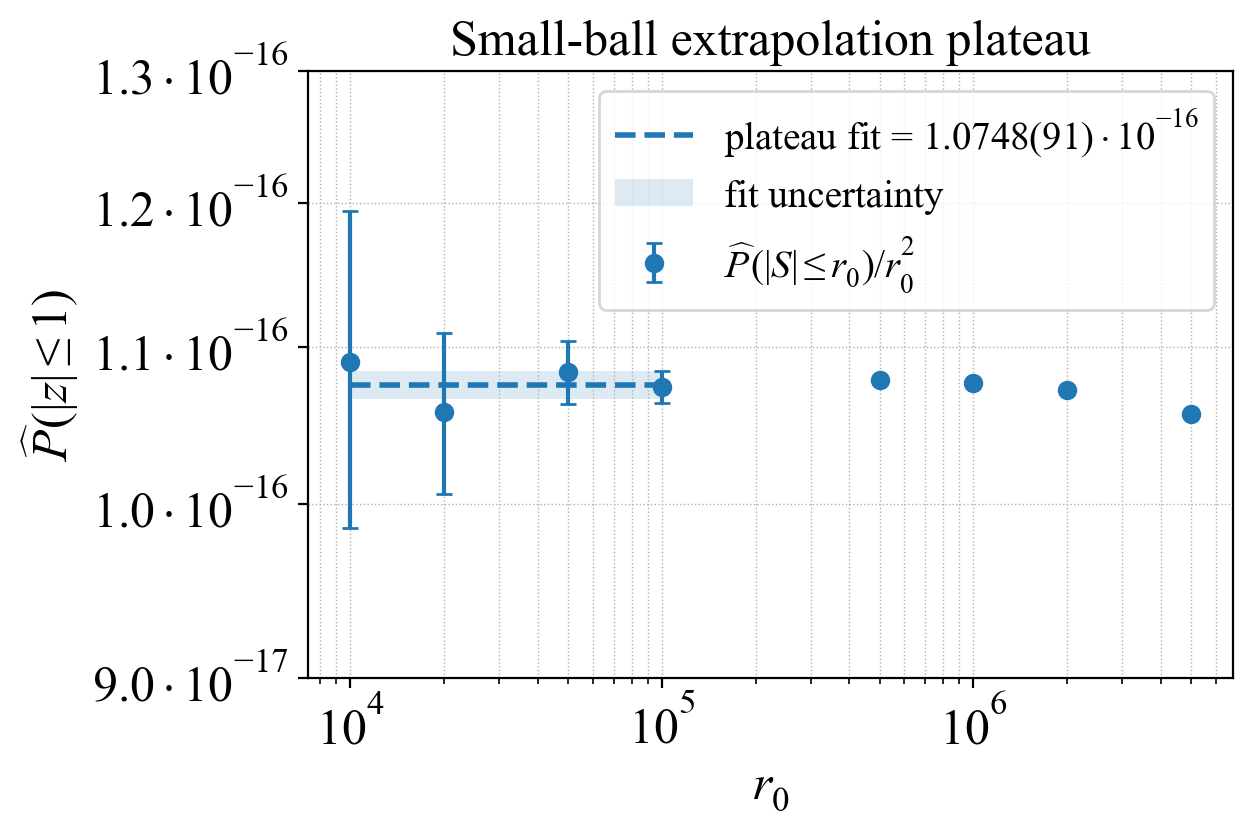}
\caption{Empirical determinations of $P(\az \leq 1)$ from Monte Carlo evaluations of $P(\az \leq r_0)$, with $r_0$ values in the ``intermediate'' range $1 \ll r_0 \ll \az_{\rm peak}$ corresponding to the onset of $r^2$ scaling. See text and \reftab{tab:small_ball} for further details.}
\label{fig:small_ball}
\end{center}
\end{figure}
\begin{table}[t!]
\centering
\renewcommand{\arraystretch}{1.25} 
\setlength{\tabcolsep}{8pt}         
\begin{tabular}{c c c c}
\hline
$r_0$ & hits & $P(|z|\le r_0)$ & $\widehat P(|z|\le 1)\pm\delta$ \\
\hline
$1.0\cdot 10^{4}$ & 109 & $1.090\cdot 10^{-8}$ & $1.09(10)\cdot 10^{-16}$ \\
$2.0\cdot 10^{4}$ & 423 & $4.230\cdot 10^{-8}$ & $1.058(51)\cdot 10^{-16}$ \\
$5.0\cdot 10^{4}$ & 2708 & $2.708\cdot 10^{-7}$ & $1.083(21)\cdot 10^{-16}$ \\
$1.0\cdot 10^{5}$ & 10733 & $1.073\cdot 10^{-6}$ & $1.073(10)\cdot 10^{-16}$ \\
$5.0\cdot 10^{5}$ & 269426 & $2.694\cdot 10^{-5}$ & $1.0777(21)\cdot 10^{-16}$ \\
$1.0\cdot 10^{6}$ & 1076190 & $1.076\cdot 10^{-4}$ & $1.0762(10)\cdot 10^{-16}$ \\
$2.0\cdot 10^{6}$ & 4286966 & $4.287\cdot 10^{-4}$ & $1.07174(52)\cdot 10^{-16}$ \\
$5.0\cdot 10^{6}$ & 26401016 & $2.640\cdot 10^{-3}$ & $1.05604(21)\cdot 10^{-16}$ \\
\hline
\end{tabular}
\caption{Small-ball probabilities from a Monte Carlo scan with $N=1\cdot 10^{10}$ samples and $c_i\sim\mathrm{Unif}[-5.0,5.0]$. We report the raw hit count $\#\{|z|\le r_0\}$, the empirical $P(|z|\le r_0)$, and the extrapolated $\widehat P(|z|\le 1)=P(|z|\le r_0)/r_0^2$ with binomial uncertainty propagated.}
\label{tab:small_ball}
\end{table}
Consequently, obtaining {\em one single} instance with $\az \sim O(1)$ would require a scan of order $10^{15}$ points, far beyond any practical brute-force exploration of the parameter space.

\medskip

\noi {\em Geometric interpretation.} Geometrically, the complex vector $\adotc$ typically populates a region of radius $\sim 10^{7}$ in the complex plane. Reaching $\az \sim O(1)$ corresponds to forcing this vector to lie inside a disk whose radius is smaller by seven orders of magnitude. Such configurations therefore occupy a vanishingly small fraction of the full 35-dimensional parameter space and require highly correlated cancellations among many independent contributions. Two simple examples of such cancellations will be explored in \refsec{app:fa:cases_i_and_ii}.

These considerations show that the weak-amplitude regime $\az \sim O(1)$, while not
excluded mathematically, is geometrically exceptional.
This supports the interpretation of the bound \refeq{eq:fa_bound} as a conservative absolute lower limit on $f_a$: in the overwhelming majority of parameter space, the implied constraint is stronger by many orders of magnitude.

\subsection{On specific configurations yielding weak-amplitude dominance}\label{app:fa:cases_i_and_ii}

In this subsection we examine quantitatively whether the bound \refeq{eq:fa_bound} can be approached in special corners of parameter space. We analyze two representative situations that can lead to $\azexpl \sim O(1)$: {\em (i)} dominance by a single coupling, and {\em (ii)} engineered cancellations between one coupling and the sum of the rest. We consider them in turn.

\subsubsection{Case (i): single-coupling dominance}
\label{app:fa:case_i}

\noi We first consider the situation in which only a single UV coupling $c_k$ is varied, while all other $c_{j\neq k}$ are set to zero. In this case,
$|z| = |1 + \alpha_k c_k|$ (no sum over $k$), and the bound on $f_a$ depends only on $c_k \in [-5,5]$. Except in a small neighborhood of $c_k \simeq -1/\alpha_k$, one finds that $|1+\alpha_k c_k|$ is of order unity or larger, so that the resulting bound on $f_a$ is equal to or stronger than $\fab$.

\begin{figure}[t!]
\begin{center}
\includegraphics[width=0.92\textwidth]{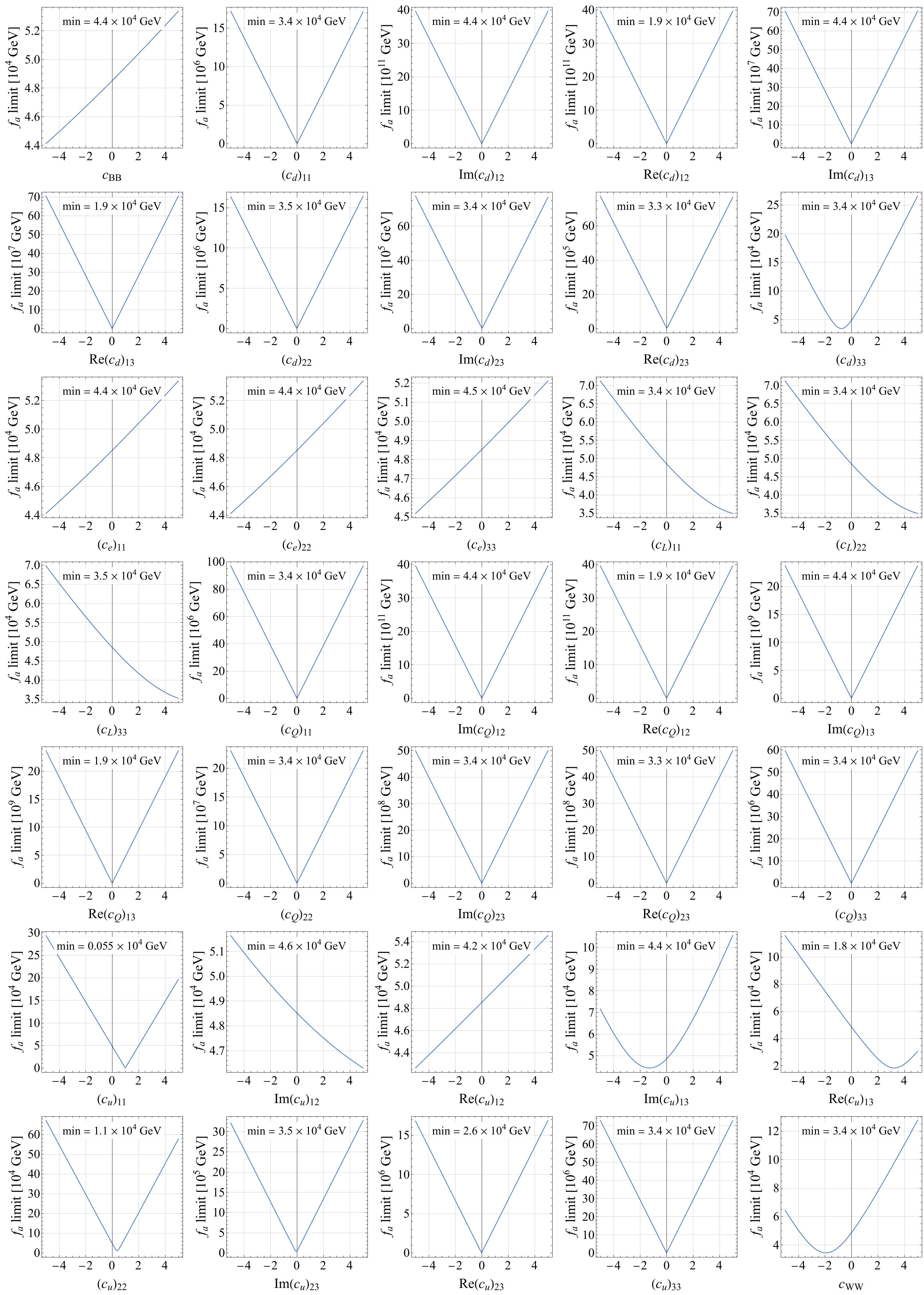}
\caption{$f_a$ limit (\refeq{eq:faz}) as a function of one single parameter at a time in the range $[-5,5]$. The inset shows the minimum $f_a$ limit within this range.}
\label{fig:faPlotGrid}
\end{center}
\end{figure}

A weaker bound can arise if $c_k$ is tuned to cancel the constant term, namely $c_k \simeq -1/\alpha_k$, which produces a cusp-shaped minimum in $f_a(c_k)$. This behaviour is illustrated in \reffig{fig:faPlotGrid}, where the minimum of the bound occurs at $c_k=-1/\alpha_k$ whenever this value lies within $[-5,5]$. In some cases the minimum of $f_a$ falls slightly below $\fab$, but only for comparatively large couplings (magnitude $\sim 2$) and only when the mentioned cancellation condition is enforced.

Such a cancellation requires a specific numerical relation between two quantities that are a priori unrelated: the UV coupling $c_k$, defined at the PQ scale, and the coefficient $\alpha_k$, which encodes renormalization-group evolution over many decades in energy. As no known symmetry or dynamical argument correlates these quantities, achieving $c_k \simeq -1/\alpha_k$ amounts to fine tuning. We therefore conclude that single-parameter dominance does not generically evade the bound \refeq{eq:fa_bound}.

\subsubsection{Case (ii): cancellations between one parameter and the sum of the rest} \label{app:fa:case_ii}

\noi We now consider the possibility that one UV coupling $c_i$ cancels the remaining contributions to $z=1+\boldsymbol{\alpha}\cdot\mathbf{c}$, leading to $|z|\sim O(1)$. This requires
\be
c_i \simeq - \frac{1}{\alpha_i}\left(1 + \sum_{j\neq i}\alpha_j c_j \right) .
\label{eq:ci_sum_cj_app}
\ee
The 35 independent parameters $c_i$ comprise the diagonal gauge and fermion couplings as well as the real and imaginary parts of off-diagonal fermion couplings; in all cases $c_i$ is real by construction. The right-hand side of \refeq{eq:ci_sum_cj_app}, however, is in general complex. As a result, \refeq{eq:ci_sum_cj_app} amounts to two independent conditions: one fixes $c_i$ to the real part of the right-hand side, while the other enforces the vanishing of its imaginary part.

To see when the first condition can be satisfied without additional tuning, rewrite the cancellation condition as
$$
\alpha_i c_i \simeq -\sum_{j\neq i}\alpha_j c_j~.
$$
If $|\alpha_i|$ is among the largest coefficients, the ratios $|\alpha_j/\alpha_i|$ are at most $O(1)$, and the equality can be fulfilled with generic values of the coefficients $c_j$. These parameters represent UV couplings and are therefore expected to remain perturbative, i.e. not parametrically larger than unity.
By contrast, if $|\alpha_i|$ is not among the largest, the sum necessarily contains terms with $|\alpha_j/\alpha_i|\gg 1$. Under the same perturbativity requirement, the $c_j$ cannot compensate such enhanced contributions individually; satisfying the condition then requires correlated cancellations among several $c_j$.

Even if the real-part condition is satisfied, the additional requirement that the imaginary part vanish imposes an independent constraint. It demands correlated choices among several a priori unrelated complex contributions. These involve unrelated phases and display a wide hierarchy in magnitude, both features originating from the coefficients $\alpha_j$, which arise from RG evolution over many decades in energy and thereby link IR and UV dynamics. Given the absence of any symmetry relating them, such an imaginary-part cancellation is strongly non-generic.

\begin{figure}[t!]

\begin{center}
\includegraphics[width=0.98\textwidth]{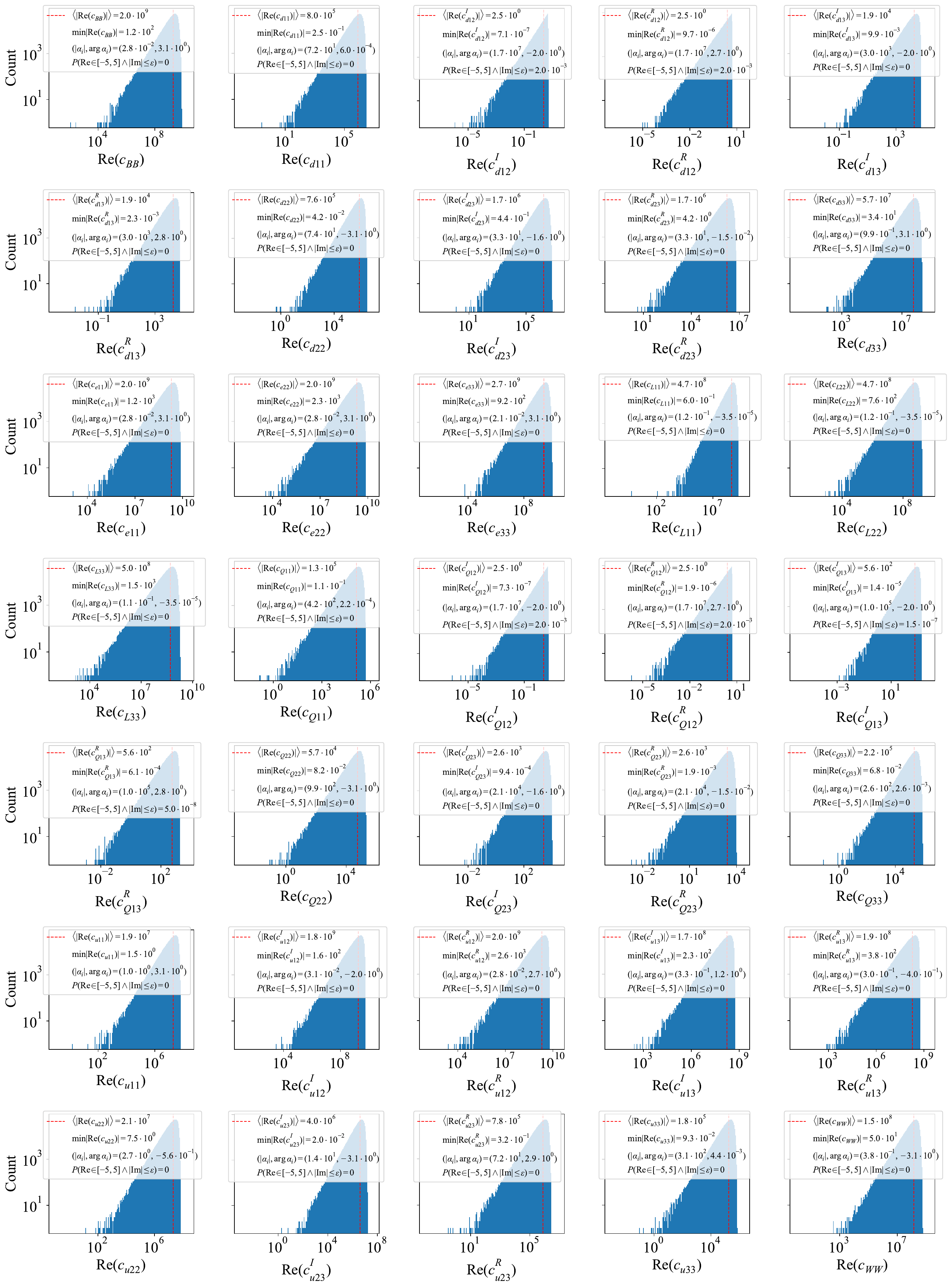}
\caption{Distributions of the real part of \refeq{eq:ci_sum_cj_app}, for all possible choices of $c_i$. See text for details on the legend entries.}
\label{fig:ci_cj_Re}
\end{center}

\end{figure}

These statements are illustrated in \reffigs{fig:ci_cj_Re}{fig:ci_cj_Im}, which display the distributions of the real and imaginary parts, respectively, of the right-hand side of \refeq{eq:ci_sum_cj_app} for each of the 35 possible choices of $c_i$. In the plots, the parameters corresponding to the real and imaginary parts of off-diagonal fermion couplings, $\Re(c_{Q,u,d})_{ij}$ and $\Im(c_{Q,u,d})_{ij}$, are abbreviated as $c^{R,I}_{Q,u,d\,ij}$ to avoid clutter. These denote independent real scan parameters; the $\Re$ and $\Im$ operations shown in the figures always refer to the right-hand side of \refeq{eq:ci_sum_cj_app}. 
The results are obtained from a Monte Carlo scan with $2\times 10^7$ random samples per parameter.

In each panel, the red dashed line indicates the average absolute value of the distribution, which characterizes the typical scale generated by random UV coefficients; its numerical value appears as the first legend entry. The second legend entry reports the smallest absolute value obtained for that parameter as a result of scanning over the remaining 34 couplings. The third legend entry identifies the corresponding $\alpha_i$ coefficient (see \reftab{tab:aici}), making explicit which parameters are associated with the largest RG-induced weights.

Only the four parameters corresponding to the largest $|\alpha_i|$ admit support for the real-part condition within the allowed interval $[-5,5]$. Even in these cases, the imaginary-part distribution remains strongly peaked away from zero. The last legend entry reports the fraction of scan points satisfying both conditions simultaneously within a numerical tolerance $\epsilon=0.01$. This fraction is found to be zero or statistically negligible in all cases. (The tolerance $\epsilon=0.01$ is chosen to be small compared to the typical scale of the distributions; tightening it further only decreases the already negligible fraction of satisfying points within the explored scan size.)

We conclude that multi-parameter cancellations capable of yielding $|z|\sim O(1)$ are not excluded mathematically, but are geometrically and statistically implausible. This reinforces the interpretation of the bound \refeq{eq:fa_bound} as a general conservative lower limit on $f_a$.

\begin{figure}[t!]
\begin{center}
\includegraphics[width=0.98\textwidth]{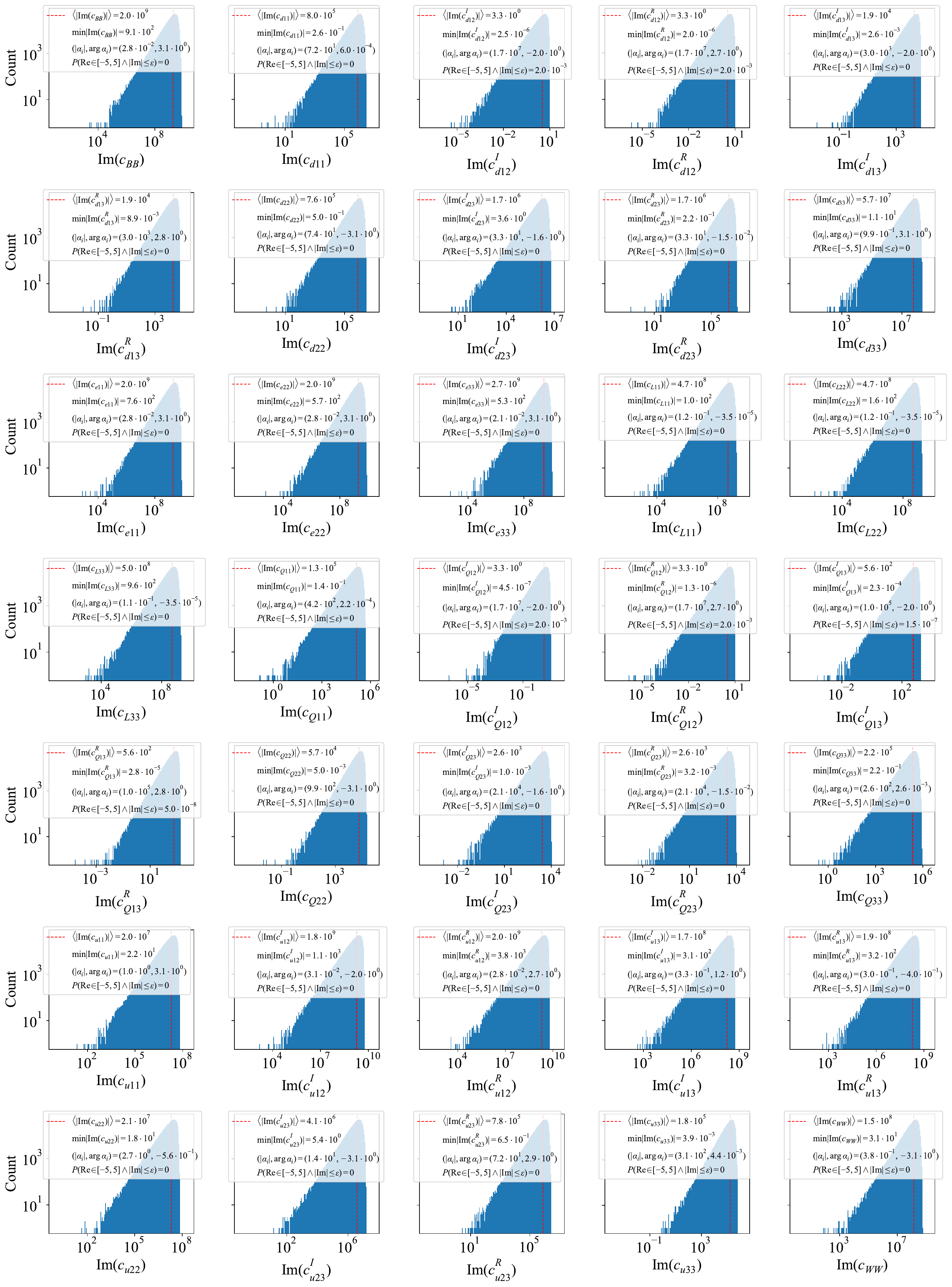}
\caption{Distributions of the imaginary part of \refeq{eq:ci_sum_cj_app}, for all possible choices of $c_i$. See text for details on the legend entries.}
\label{fig:ci_cj_Im}
\end{center}

\end{figure}

\end{document}